# Testability Refactoring in Pull Requests: Patterns and Trends

Pavel Reich
*Applied Software Technology*
*Universität Hamburg*
Hamburg, Germany
pavel.reich@studium.uni-hamburg.de

Walid Maalej
*Applied Software Technology*
*Universität Hamburg*
Hamburg, Germany
walid.maalej@uni-hamburg.de

*Abstract*—To create unit tests, it may be necessary to refactor the production code, e.g. by widening access to specific methods or by decomposing classes into smaller units that are easier to test independently. We report on an extensive study to understand such composite refactoring procedures for the purpose of improving testability. We collected and studied 346,841 java pull requests from 621 GitHub projects. First, we compared the atomic refactorings in two populations: pull requests with changed test-pairs (i.e. with co-changes in production and test code and thus potentially including testability refactoring) and pull requests without test-pairs. We found significantly more atomic refactorings in test-pairs pull requests, such as Change Variable Type Operation or Change Parameter Type.

Second, we manually analyzed the code changes of 200 pull requests, where developers explicitly mention the terms "testability" or "refactor + test". We identified ten composite refactoring procedures for the purpose of testability, which we call testability refactoring patterns. Third, we manually analyzed additional 524 test-pairs pull requests: both randomly selected and where we assumed to find testability refactorings, e.g. in pull requests about dependency or concurrency issues. About 25% of all analyzed pull requests actually included testability refactoring patterns. The most frequent were *extract a method for override* or *for invocation*, *widen access to a method for invocation*, and *extract a class for invocation*. We also report on frequent atomic refactorings which co-occur with the patterns and discuss the implications of our findings for research, practice, and education.

## I. INTRODUCTION

Developers spend a significant proportion of their time on testing and quality assurance [1]. In recent years, automated testing has become popular [2], in particular unit testing, where a software unit is tested separately from the entire system. This typically includes instantiating a unit, invoking methods in the unit under test, and observing its state or behavior using assertions [3]. If not designed with unit testing in mind, software systems might need extra refactoring effort to improve its observability and controllability [4], enabling a better testability. For instance, a dependency of a class under test might need to be replaced with a mock object to simulate the behavior of the dependency [5]. One such example can be found on GitHub in the pull request number 1449 of the pentaho-platform project as shown in Figure 1. A separate method `getXmlaExtra` is extracted from the production code (top part), in order to override it in the test code to return a mock object (bottom part). Such a co-change of the production and test code in a single pull request is obviously intentional for the purpose of improving unit testability.

Fig. 1: Example of a test-pair PR where the production code (top part) is refactored to extract a method for using it in the test code (bottom part).

To understand how developers refactor production code in order to enable and ease its unit testing, we conducted an extensive study using a mix of manual and automated analyses. The goal of the study is to identify and characterize common *testability refactoring patterns* in open source projects. We automatically mined thousands and manually analyzed 724 of pull requests (PR) from 621 Java projects hosted on GitHub. PRs represent well-defined units of changes and contain one or more commits with the changed files as well as meta information and text describing the changes. While PRs often contribute new functionality or bug-fixes, there are also cases where production and test code are changed at once [6]. We focus our study on these cases and call them **Test-Pairs PRs**. Two such PRs are jmxtrans/171 with refactoring to improve test code coverage and kie-wb-common/367 with a new unit test and corresponding refactoring in production code.

The contribution of this paper is twofold. First, we report on differences between Test-Pairs PRs and other PRs regarding the code changes. We found that Test-Pairs PRs are more refactoring intensive than other PRs, with specific frequent refactorings such as changing variable type, changing parameter or return type, extracting an operation, and adding a parameter (see Figure 3). These findings call for further research on understanding and supporting the test-oriented refactoring in pull-based development. Second, we identify 10 recurring higher-level refactoring patterns implemented in Test-Pairs PRs to improve testability. These include, e.g., *extract method for override* or *add constructor parameter* (see Table III for the full list).

Most patterns aim at breaking certain dependencies between the classes under test to enable testing them in separation. Our patterns can be used as refactoring templates in IDEs, as catalog for education and training, or as testability guidelines for contributors in open source projects. Our results also indicate the pattern frequencies, in which pull requests they can be expected, and with what atomic refactorings they co-occur. Our analysis pave the way towards (automatically) learning how higher-level refactoring is composed of atomic minable refactorings. We share our replication package and the catalog of patterns[1] and discuss how it can be used in academia and industry.

The remainder of the paper is structured as follows. Section II introduces our research questions, method, and data. Then, Section III reports on the automated analysis of test-pairs PRs. Section IV introduces the identified catalog of testability refactoring patterns, while Section V reports on their prevalence and characteristics. Finally, Section VI discusses our findings, Section VII summarizes related work, while Section VIII concludes the paper.

## II. RESEARCH METHODOLOGY

### A. Research Questions and Terminology

There are multiple definitions for software testability as discussed by Garousi et al. [7]. In this work we focus on unit-testability. We thus refer to **testability** as the degree to which a software unit or a component can be tested in isolation [8]. Our overall goal is to better understand testability in open source setting, in particular how and how often developers refactor production code in order to develop unit-tests for it.

Generally, developers refactor code to make it more maintainable, reduce code smells, or for other design reasons that do not change the software behavior, including improving the testability [9]. Fowler [10] suggested a catalog of common fine-grained refactoring actions such as changing return type or extracting an interface, which can automatically be detected using repository mining tools such as RefactoringMiner [11]. We refer to those as **atomic refactorings** or simply **refactorings**. First, we aim at mining and analyzing those refactorings in context of test-pair changes. Second, we seek for meaningful, higher-level, composite refactoring patterns that include coherent refactorings conducted for the purpose of testability. We call those patterns **testability refactoring patterns**. We aim at exploring where and how often to expect these patterns and what code changes they typically include. As a result, this paper focuses on the following research questions:

- **RQ1**: Are there particular atomic refactorings trends that characterize Test-Pairs PRs?
- **RQ2**: What recurrent higher-level refactoring patterns do developers perform for the goal of improving testability?
- **RQ3**: How prevalent are testability refactoring patterns and what characterizes the corresponding code changes?

[1] https://github.com/icse2023preich/testability-refactoring-patterns

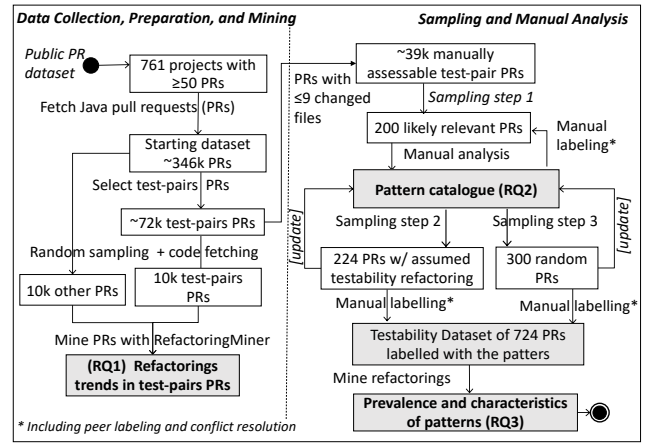

Fig. 2: Research overview: steps and corresponding datasets.

### B. Data Collection, Preparation, and Mining

To answer the research questions, we fetched open source projects hosted on GitHub and mined their pull requests, consisting of one or more commits (i.e. the actual code change) as well as the meta information about the contribution (i.e. a mandatory title, optional description, and other elements). Figure 2 overviews the main steps we followed to collect, prepare, and analyze the data.

In order to collect active GitHub projects, we used a recent public dataset [12] with 3,347,937 pull requests in 1,823 Java projects, from which we selected 761 projects each with at least 50 pull requests. This ensured a minimum level of project maturity and popularity of PRs. From the set of 761 unique projects, we fetched 508,240 corresponding PRs using GitHub GraphQL V4 API in JSON format with pagination for every 100 items including the following fields: url, bodyText as description of the PR, title, total list of files, derived count of added/removed lines for every file in the PR. From this set of PRs, we selected 346,841 PRs with at least one changed .java file. This represents our starting dataset in Figure 2.

We then split this dataset into PRs with at least one test pair and PRs without any test-pairs. A source-test pair represents a java class under test (MyClass) with the corresponding unit-test (MyClassTest) linked using suffix-based naming convention [13]. This led to 72,115 Test-Pairs PRs and 274,726 other PRs, which represent the basis for answering RQ1.

From this basis, and since the automated retrieval of the source code corresponding the to PRs is restricted in GitHub,

we created a *random sample* of 10,000 Test-Pairs PRs and 10,000 other PRs, which we use in our automated analysis. To mine fine-grained refactorings from PRs, we used RefactoringMiner version 2.3.2 [11]. The tool can detect 93 refactoring patterns including standard Fowler refactorings. We focused on analysing refactorings of production code and excluded refactorings applied to test code itself. Since RefactoringMiner processes all commits in a PR and in order to exclude merge/rebase commits, we excluded refactorings from commits that have more classes affected than files in the PR itself. Those would otherwise artificially inflate the frequency of refactorings not introduced in the actual PR.

Next, we selected manually assessable Test-Pairs PRs. That is, PRs with at least one test pair and at most 9 changed files. According to Miller's law, 9 is the upper number of changed files that can be well handled by average human's short-term memory [14]. This finally resulted in 39,077 PRs that satisfy those criteria.

*C. Sampling and Manual Analysis*

Since we are unaware of any public testability dataset that we can use to study testability refactoring patterns, we created in a extensive manual analysis process our own dataset, which we share in our replication package[2]. Overall, we followed a mixed method approach, including methodological elements from Grounded Theory [15], Thematic Analysis [16], and Manual Content Analysis [17]. This includes incremental sampling (to sample relevant PRs), themes identification and theory building (to identify the patterns), as well as review, refinement, labeling, and reliability checking (to annotate the PRs with the single patterns they include). The unit of analysis was a manually assessable Test-Pairs PR (with 9 changed files at most [14]), including all code changes as well as the PR details and meta-data. Both authors together with three hired and trained postgraduate students were involved in this process. The results include a list of 10 patterns and a set of 724 PRs which are annotated to (exclusively/not-exclusively) include testability refactoring or not; and if yes which of the patterns do they include. 441 of these PRs were annotated independently at least by two persons while the rest were analyzed and annotated at least twice by the first author.

To create this manually labeled dataset, we encountered two major **challenges**. First, we had to balance between the reliability of the analysis and the required manual effort. Analysing code changes of PRs not only requires advance development skills, but is also a cognitive demanding, time-consuming task. One assessment of a single PR required 10 or more minutes, sometimes much longer, e.g. in cases which needed discussion to resolve labeling conflicts. The second major challenge was the sampling. Since we were looking for a rather rare phenomenon without prior empirical evidence, a fully random dataset was very likely to not lead to meaningful results and we would certainly have missed relevant patterns this way. We thus followed a targeted incremental sampling over three major steps, starting from our assumptions of where we could find testability refactoring based on specific keywords. We then used randomization whenever a subsample was too large to be reviewed completely. Table I shows the sampling steps, corresponding search keywords (i.e. masks), and the final counts of retrieved and manually reviewed PRs.

[2]https://github.com/icse2023preich/testability-refactoring-patterns/blob/main/README.md

TABLE I: Sampling steps and masks with the corresponding counts of retrieved (Ret.) and manually reviewed (Rev.) PRs.

| Sampling Step | ID | Mask | Keywords sought for | Ret. | Rev. |
|---|---|---|---|---|---|
| 1 | M1 | testability_body | testability OR testable | 109 | 109 |
| 1 | M2 | testability | testability OR testable | 18 | 18 |
| 1 | M3 | Refactor for test | refactor AND (test OR junit) | 73 | 73 |
| 2 | M4 | Dependency | depend | 264 | 50 |
| 2 | M5 | Concurrency | concurren OR thread OR sleep OR latch | 459 | 50 |
| 2 | M6 | Network | network OR socket OR connectivity OR connection | 221 | 50 |
| 2 | M7 | Singleton | singleton | 24 | 24 |
| 2 | M8 | Inject | inject OR wire OR wiring | 143 | 50 |
| 3 | M9 | test | test OR junit | 2,268 | 150 |
| 3 | M10 | Other | NONE | 35,633 | 150 |
|  |  | Total |  | 39,212 | 724 |

In **sampling step 1**, we looked for likely relevant PRs, where the terms "testablity" or "testable" are explicitly mentioned in the body (M1) or the title (M2). We identified 109 and respectively 18 of such PRs and manually analyzed all of them. Other likely relevant PRs have titles including both terms "refactor" + "test" or "refactor" + "junit". We found 73 such PRs (M3) and analyzed all of them too. In sum, we started with 200 PRs (M1+M2+M3) which are likely relevant for testability refactoring.

The first author who has over decade of Java development experience read each PR text and reviewed its code changes and categorized the PR into a) irrelevant, b) PR only covers refactoring for testability or c) PR includes refactoring for testability but not exclusively. PRs were presented in a Jupyter notebook interactive table with a clickable URL to the PR and a drop-down containing 4 values: unseen, irrelevant, only_ref_for_test and incl_ref_for_test. Candidate patterns were noted separately with examples of their occurrences. The intermediate results were then discussed in detail with the second author and a small random sample of 10PRs from the reviewed PRs were rechecked. This led to the first set of testability refactoring patterns (the first 9 listed in Table III). With this set, the first author went back to the 200 PRs and labeled presence of one or more patterns.

With this first stable list of patterns, we moved to analyze additional PRs. After observing that PR body is generally noisy with keywords not directly characterizing the PR, we searched only the titles for additional keywords in the remaining steps. In **sampling step 2**, we retrieved potentially relevant PRs, which refer to architectural situations that might be hard to test and might require refactoring to enable or improve testability. Based on the literature on dependency-breaking techniques (as described in by Feather [18]) and based on our experience, this might be the case where:

- The Test-Pairs PR handles a dependency (M4)
- Concurrency or multi-threading are involved (M5)
- Sockets or network operations are used (M6)
- The Test-Pairs PR is about singleton (M7) [19] [20]
- Dependency injection or wiring are involved (M8)

These potentially relevant PRs correspond to M4-M8 in Table I. For each of these cases, we randomly sampled from the retrieved PRs up to 50, which we manually analyzed as described above. This step resulted into 224 additional PRs.

The goal of **sampling step 3** was to ensure that we do not completely miss important and frequent refactorings through the masking and to get an indication (through random sampling) about the overall patterns' frequency in Test-Pairs PR. We thus retrieved and reviewed 150 random PRs that mention test in the title (i.e. about testing) (M9) and another 150 random PRs without any restrictions (M10).

To check and ensure the reliability of our manual analysis we additionally hired three graduate students with several years experience of Java development to independently re-review and re-label a random subset (at least 50%) of our entire sample. Those additional coders followed the same procedure described above (first review if the PR is relevant, then label the patterns). They also used a concrete guidelines (coding guide in the replication package), which describe the labeling task with positive and negative examples. We trained the three coders by explaining the task, examples and asking them to label a small set of 10 PRs which we discussed and corrected with them to get familiar with the task. Finally, 441 PRs were independently analyzed at least by two persons. Conflicted cases were discussed with the second author who also overviewed the whole process, assignments, reviewer briefing, and consolidation of the patterns after the iterations. Most disagreements were due to one of the coders missing one of the patterns, as we discuss in Section VI.

The additional analysis after sampling steps 2 and 3 and the additional pair-coding led to the 10th pattern in Table III as well as 10 other distinct refactoring procedures which were clearly for the goal of testabiltiy but which we were unable to give a general label.

### III. MINING TEST-PAIRS PULL REQUESTS (RQ1)

We report on the analysis results concerning the types and frequency of the atomic refactorings in production code: when developers change production and test code together i.e. Test-Pairs PRs, vs. other PRs. Table II shows an overview of the analyzed sample, 10,000 test-pairs PRs vs. 10,000 other PRs without test-pairs. In total, RefactoringMiner identified 273,201 refactorings in 7,782 (61.7%) Test-Pairs PRs and 114,427 refactorings in 4,839 (38.3%) other PRs. On average, Test-Pairs PRs had significantly more refactorings (i.e. 34.9 refactorings per PR) against 23.6 refactorings per PR in the other PRs. Test-Pairs PRs had also more added and deleted lines of code: on average, 670 changed lines of code in Test-Pairs PRs against 225.9 LOC in other PRs. This indicates that Test-Pairs PRs are *more refactoring-intensive* than other PRs. It is also important to note, that Test-Pairs PRs seem to be overall larger with higher numbers of churn LOC, added LOC, and deleted LOC. To understand whether the higher number of refactoring has led to the higher numbers of changes in LOC or vice-versa would require further analysis in a controlled environment.

TABLE II: Overview of the two mined PR datasets.

| Metric | Test-Pairs PRs | Other PRs |
|---|---|---|
| PR count | 10,000 | 10,000 |
| PRs with refactorings | 7,822 | 4,839 |
| Mined refactorings | 273,201 | 114,427 |
| Average LOC added per PR | 483.9 | 157.8 |
| Average LOC deleted per PR | 191.8 | 70.7 |
| Average churn LOC per PR | 675.6 | 228.6 |
| Average mined refactorings per PR | 34.9 | 23.6 |

When comparing the single atomic refactorings between the samples, we also observed several interesting trends, shown on Figure 3. In general, we found that the distribution of refactorings to be rather different between the two types of PRs. Add Attribute Annotation is roughly 2x more frequent in Test-Pairs PRs than in other PRs. Refactorings such as Add Attribute Annotation, Move Class, Rename Method, Add and Change Parameters as well as Change Return Type are more frequent in Test-Pairs PRs. Change Variable Type seems to be overall the most frequent refactoring with 3 to 4 occurrences in each Test-Pairs PR. The other refactorings (in particular those concerning variables and operations) seem to have a similar prevalence in both samples.

Overall, we observed that the trends within the samples are rather stable. The variation between the single PRs seem rather low, except for Add Method Annotation in the Others PRs. This particular variation might result from different usage policies of annotations in the different projects.

### IV. TESTABILITY REFACTORING PATTERNS (RQ2)

We identified 10 recurrent patterns, which we split into five groups based on the targeted code element. As Table III shows, we found 230 occurrences of the patterns in 184 PRs.

TABLE III: Testability refactoring patterns and their counts in the 724 manually analyzed PRs.

| ID | Pattern name | Count | % |
|---|---|---|---|
| 1 | extract_method_for_override | 51 | 22.2 |
| 2 | extract_method_for_invocation | 39 | 17.0 |
| 3 | widen_access_for_invocation | 35 | 15.2 |
| 4 | extract_class_for_invocation | 29 | 12.6 |
| 5 | add_constructor_param | 25 | 10.9 |
| 6 | extract_class_for_override | 15 | 6.5 |
| 7 | create_constructor | 10 | 4.3 |
| 8 | widen_access_for_override | 9 | 3.9 |
| 9 | override_system_time | 4 | 1.7 |
| 10 | extract_attribute_for_assertion | 3 | 1.3 |
|  | other | 10 | 4.3 |
|  | Total | 230 | 100.0 |

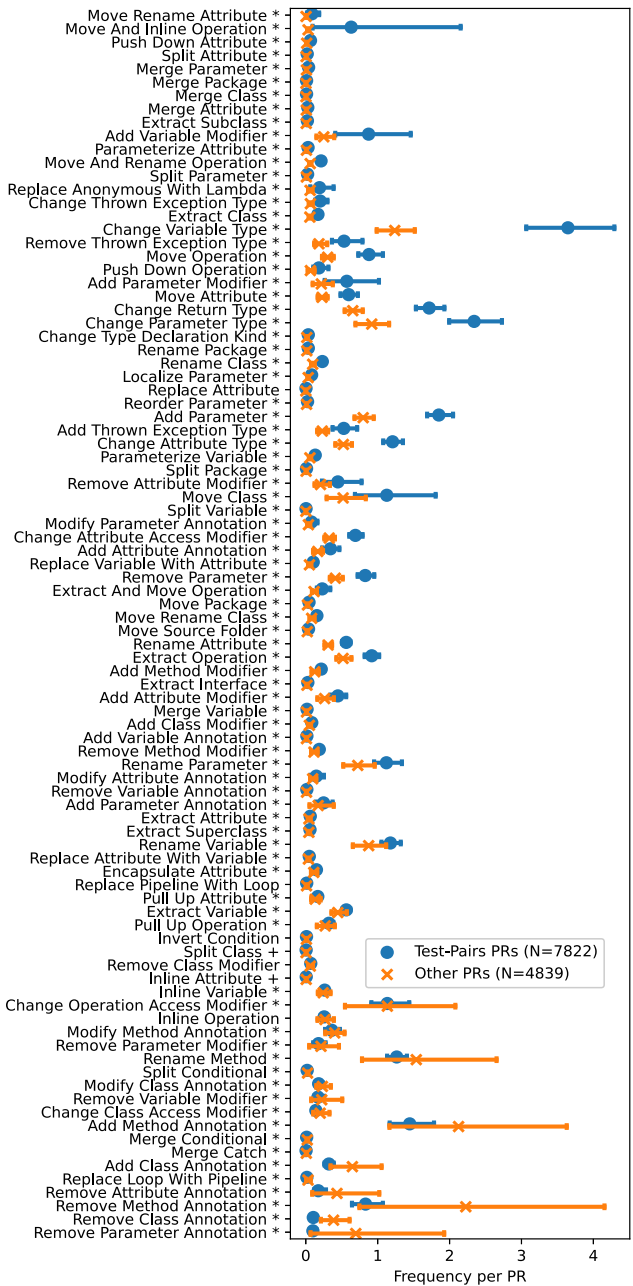

Fig. 3: Frequency of atomic refactorings for Test-Pairs PR and Other PRs. Refactorings with significant differences in frequency (p-value < 0.05) are marked with *.

## A. Extract Method

The most frequent testability refactoring pattern in our dataset is *extract method*. It was used in about 39.1% of occurrences, with two main goals: to override the behavior of the class under test or to test it separately.

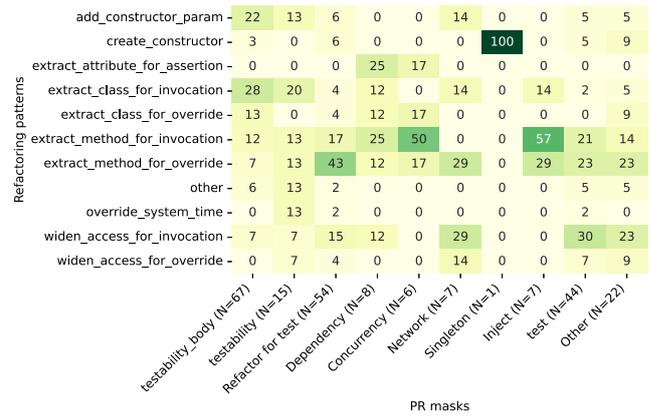

Fig. 4: Testability refactoring patterns found across PR masks.

```
LocalScheduler scheduler = Mockito.spy(LocalScheduler.class);
Process mockProcess = Mockito.mock(Process.class);
Mockito.doReturn(mockProcess).
    when(scheduler).
    startExecutorProcess(Mockito.anyInt());
```

Fig. 5: Override method result with a mock using Mockito.spy in test incubator-heron/536.

**Extract method for override** In about 22.2% of analyzed refactorings for testability in our sample, methods are extracted in production code to be overridden in test code in an anonymous subclass or using Mockito framework. For example, in PR pentaho-platform/1449, the method getXmlaExtra in the class under test OlapServiceImpl has been extracted as protected, as shown on Figure 1. In the test code the method is overridden in an anonymous subclass to return a mock object.

Such methods are often declared as protected or package private to limit access to the methods from classes outside the package. Instead of subclassing the class under test, the Mockito.spy[3] technique is often used to generate a subclass using CGLIB. This technique reduces the amount of necessary code. For example, in the PR incubator-heron/536, the method LocalScheduler.startExecutorProcess is overridden using Mockito.spy, as shown in Figure 5.

This approach is useful to isolate the class under test from its dependencies, to replace real HTTP calls with provided HTTP responses (as seen in knowm/XChange/20), or to programmatically emulate an exception in a thread created by a class under test (as seen in jchambers/pushy/43). This pattern corresponds to Feathers' *Extract and Override Call*, *Subclass and Override Method* and *Extract and Override Getter* [18].

**Extract method for invocation** In 17.0% of testability refactoring pattern occurrences, developers have extracted methods to test them separately because they are considered more error-prone, contain more complex logic or to isolate from UI/network/databases. For example, in Topsoil/162, the method writeSVGToOutputStream is extracted and unit-tested separately from UI functionality. In another example azkaban/1975, the method updateExecutions is extracted and tested separately.

---

[3]https://javadoc.io/static/org.mockito/mockito-core/3.6.28/org/mockito/Mockito.html#spy-T-

In at least two cases, extract_method_for_override and extract_method_for_invocation have been mistaken by the annotators. The borderline between the two can be defined as follows: "for override" is used to override functionality of production class by subclassing or mocking, whilst "for invocation" is used to invoke the method in order to prepare or assert the state of unit under test.

*B. Widen Access to Methods*

Developers widen access to methods in classes under test to reach or override them from test code. The annotation @VisibleForTesting[4] is often added or the method commented accordingly.

**Widen access for invocation** Attributes, methods, and classes themselves can have different visibility in java: private (visible to the same class), protected (visible to subclasses), package-private (visible to members of the same package) and public (visible to all). It may be necessary to widen access to a method to invoke it from a unit-test in order to control or observe the state of the class under test. For example, in oryx/164 private method calcSilhouetteCoefficient in class SilhouetteCoefficient is made package-private in order to invoke it from the paired unit-test and assert the calculated result. Another example of a method made visible for invocation is Anki-Android/5996. Private classes can be made visible to a unit-test by widening access to package-private. For instance, the class NoOpAction in gitlab-plugin/335 is used to assert the result of a method invocation.

**Widen access for override** Access to existing members of classes or classes themselves can be widened to override behavior of the class under test. For instance, in OpenRefine/2839 access to method findValues is widened from protected to public in order to override it using Mockito in the test code.

*C. Extract Class*

A class under test can be decomposed into several classes not only to reduce coupling as the ultimate goal, but also to test it separately, or to isolate it from the dependencies.

**Extract class for invocation** In 12.6% of testability related refactorings, a set of methods is extracted into a separate independent class in order to test it outside the original class. For example, in jmxtrans/291 the method CloudWatchWriter.convertToDouble extracted into class ObjectToDouble and tested in ObjectToDoubleTest, which doesn't need to know about dependencies of CloudWatchWriter.

**Extract class for override** A piece of functionality in production code can be extracted into a separate class that can be replaced with a different implementation provided in a unit-test. This approach has been used in just 6.5% of testability related refactorings and is relatively rare. In metrics/516, for instance, the interface ObjectNameFactory is extracted and a mock implementation is used in a unit-test to override the behavior.

*D. Constructors and Attributes*

Constructors are used to initialize attributes of a class or to pass concrete dependencies instead of initializing the dependencies from the outside.

**Create constructor** Especially in spring-based classes, an @Autowired[5] constructor can be created to provide dependencies from the client code, as opposed to private @Autowired attributes of the class that are harder to set directly from a unit-test. In dhis2-core/2892, SmsMessageSender a constructor with all attributes is added in order to initialize the class under test from the relevant unit-test with mocked dependencies. In another example alluxio/3818, in addition to the default constructor that hard-wires dependencies, a secondary constructor is created that allows to provide dependencies from a unit-test so that mocked or custom implementations of dependencies can be provided.

**Add parameter to constructor** An extended version of an existing constructor can be created that takes one or more additional parameters to pass dependencies. For example, in openmrs-contrib-android-client/349 a new constructor is created for PatientDashboardVitalsPresenter that accepts EncounterDAO and VisitApi dependencies to set attributes of the class. In the original constructor, the two attributes were initialised with a hard-wired implementation. In general, such constructors can be made package-private or protected and marked with @VisibleForTesting annotation to limit their usage to test-code, only as seen in incubator-heron/1889.

Both patterns correspond to Feathers' pattern *Parameterize Constructor* [18]. As a disadvantage of this method, he mentioned the possibility that the extended constructor can be used in production code and using package-private access with @VisibleForTesting helps to mitigate this disadvantage.

*E. Less Common Patterns*

The patterns described above are relatively common and used in 170 PRs (sometimes multiple times within the same PR), out of 230 PRs with refactorings for testability. The remaining patterns can be found in our dataset with manually reviewed testability-related PRs.

One of the rare patterns **Override System Time** can be found in the PR azkaban/1975, where system time is overridden in the unit test using joda-time[6] DateTimeUtils.setCurrentMillisFixed method. This pattern sometimes co-occur with Add parameter to constructor described above, where an instance of Clock is provided in constructor.

---

[4]https://guava.dev/releases/19.0/api/docs/com/google/common/annotations/VisibleForTesting.html

[5]https://docs.spring.io/spring-framework/docs/current/javadoc-api/org/springframework/beans/factory/annotation/Autowired.html

[6]https://www.joda.org/joda-time/apidocs/org/joda/time/DateTimeUtils.html#setCurrentMillisFixed-long-

TABLE IV: Testability refactoring patterns in the manually reviewed PRs (according to the sampling masks). ref_for_test denotes the existence of at least one refactoring for testability in the PR. Irrelevant means no patterns observed.

| Title Mask | irrelevant | only_ref_for_test | incl_ref_for_test |
|---|---|---|---|
| testability_body | 49 (45.0%) | 15 (13.8%) | 45 (41.3%) |
| testability | 6 (33.3%) | 8 (44.4%) | 4 (22.2%) |
| Refactor for test | 38 (52.1%) | 16 (21.9%) | 19 (26.0%) |
| Dependency | 43 (86.0%) | 0 (0.0%) | 7 (14.0%) |
| Concurrency | 44 (88.0%) | 1 (2.0%) | 5 (10.0%) |
| Network | 45 (90.0%) | 1 (2.0%) | 4 (8.0%) |
| Singleton | 23 (95.8%) | 1 (4.2%) | 0 (0.0%) |
| Inject | 44 (88.0%) | 0 (0.0%) | 6 (12.0%) |
| test | 117 (78.5%) | 13 (8.7%) | 19 (12.8%) |
| Other | 131 (87.3%) | 1 (0.7%) | 18 (12.0%) |
| Total (N=724) | 540 (74.7%) | 56 (7.7%) | 127 (17.6%) |

Another rare pattern is **Extract Attribute For Assertion**, where a getter is extracted for an internal attribute in order to invoke it for assertion. For example in PR druid/2878 a @VisibleForTesting package-private getter is used to read workersWithUnacknowledgedTask attribute in a unit-test.

Finally, we found at least 10 PRs with refactorings for testability that we were unable to describe as an obvious pattern, but that were relevant for improving the testability of the production code.

## V. PREVALENCE AND CHARACTERISTICS OF TESTABILITY REFACTORING (RQ3)

Overall, we found that about 25% of manually reviewed PRs contain changes which represent one or more refactoring patterns for testability, while the other 75% of the reviewed PRs were irrelevant. Table IV shows the count and percentage of corresponding relevant PRs in our sample: in the second column those that were created only for the purpose of testability refactoring (7.73%), and in the third column those that included testability refactoring among other changes. Our results show that testablilty refactoring are found across all subsamples – also in the random subsample "Other" without any masks. Test-Pairs PRs mentioning the word testability in their body or title are likely to include at least one refactoring testability pattern (55.05% for body and 66.67% for title). Test-Pairs PRs referring to dependency, concurrency, network, singleton, and injections do not seem to be particularly relevant for testability refactoring.

We also mined the atomic refactorings in the analyzed PRs (see Figure 7) and counted lines of code added/removed in corresponding production and test code. Figure 6 shows that testability irrelevant PRs have significantly less test code added than PRs with refactorings for testability. Similarly, testability related PRs have more lines of code removed in production code: refactoring production code usually involves adding new code and removing old unused code. From 184 manually assessed testability related PRs, only three have no production code removed. For example, in alluxio/3818 a new constructor was created. We observed that refactoring for testability usually coincide with removed lines in production code that are replaced with invocations of newly extracted methods/classes. Unit-tests also often contain anonymous or named subclasses with overridden methods to modify the behavior of production code or capture parameters. Often such overrides and assertions are done using Mockito[7].

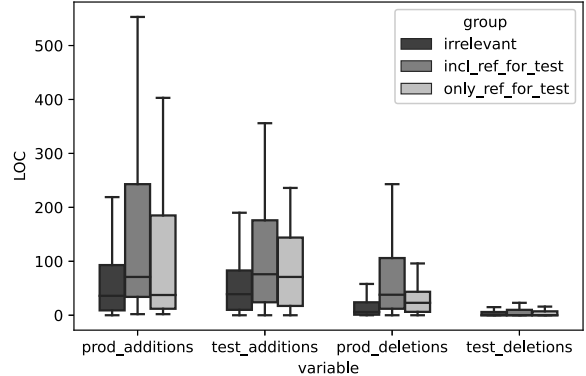

Fig. 6: Distribution of LOC by PR group.

The testability refactoring patterns can be considered as more complex and may consist of multiple Fowler's refactorings. For example, in jmxtrans/171 the class JmxResultProcessor was extracted in order to unit-test it using JmxResultProcessorTest. But from Fowler's point of view, multiple refactorings were used together: getResults, getNewResultObject and processTabularDataSupport. Methods were moved (Move Operation in RefactoringMiner's terms), the parameters MBeanInfo, ObjectInstance and Query were removed (Remove Parameter) and the parameter res has been renamed to accumulator (Rename Parameter). On a higher level, this entire refactoring can be described as extract_class_for_invocation because the class JmxResultProcessor was extracted in order to create a unit-test.

Figure 8 shows the percentage of the atomic refactoring mined in all occurrences of the testability refactoring patterns. For instance, 27% of all atomic refactoring operations to the production code, when the pattern extract_method_for_override was observed, are Extract Operation followed by 10% Change Operation Access Modifier. The remaining refactorings identified by RefactoringMiner shown in the figure (Rename Method, Rename Variable, etc.) are likely less relevant for testability. Overall, the figure also reveals that Extract Operation, Move Operation (or Extract And Move Operation) are prevalent in all patterns, except the patterns *extract attribute for assertion* and *other*.

## VI. DISCUSSION

We first discuss the implication of our findings for research and practice and then present the potential threats to validity.

### A. Significance and Implications of Findings

Our results provide insights not only into the how, and how often, but also into **why developers refactor** production code in order to create or extend unit tests: namely to override the behavior of the class under test and to validate the behavior of a smaller unit under test. A large class can be decomposed into several classes or methods can be extracted and made visible for testing, which allows to substitute inputs or dependencies with stubbed or mocked dependencies. Following Freedman's definition of testability as controllability + observability [4], it is apparent that some of the identified refactoring procedures allow unit tests to control the state of the class under test while others make the state of the system under test visible to a unit-test. Clearly, the majority of the discovered patterns aim at breaking dependencies between the classes under test in order to test them in isolation. Feathers [18] introduced dependency-breaking patterns in his book with some of them corresponding to ours.

[7]https://site.mockito.org/

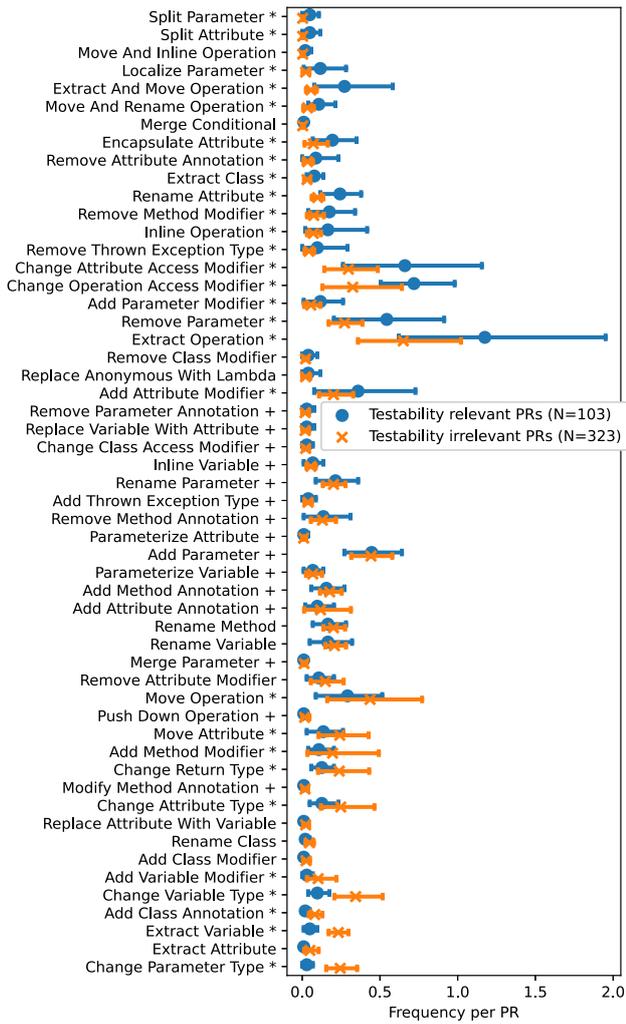

Fig. 7: Frequency of atomic refactorings per manually reviewed pull request. Comparison between testability refactoring relevant and testability refactoring irrelevant PRs.

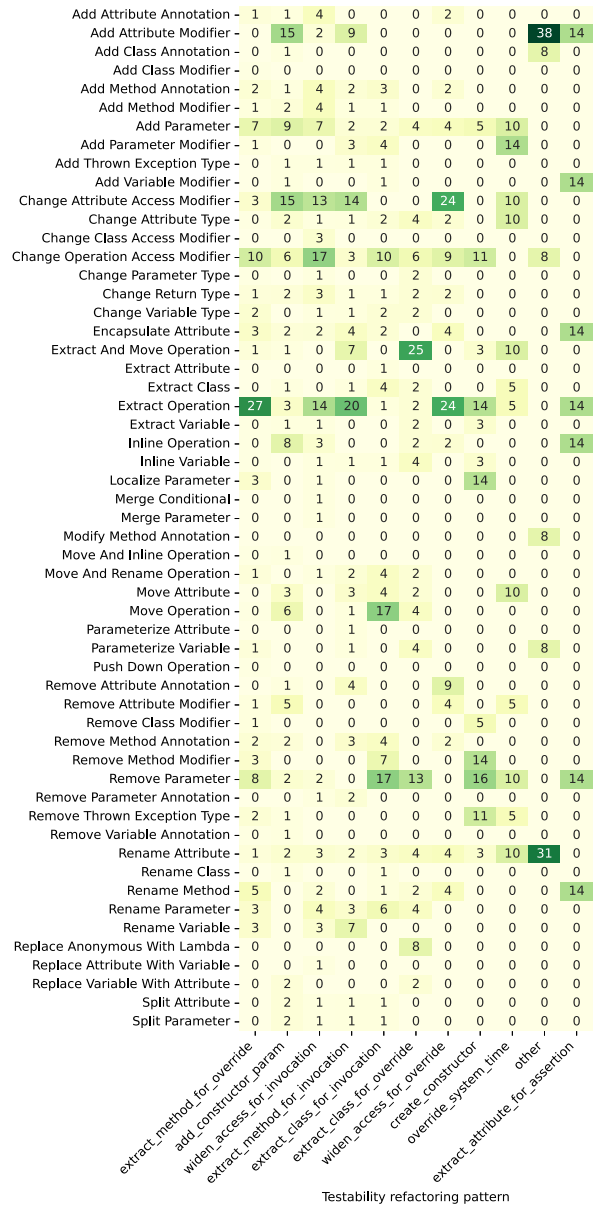

Fig. 8: Percentage of atomic refactorings in the testability refactoring patterns. Columns add to 100%.

The testability refactoring patterns, which we have identified can easily be translated into **refactoring templates** and implemented as refactoring wizards in IDEs. They guide practitioners to decompose the class under test or its methods, inject dependencies later instead of hard-wiring in the constructor, and expose methods to set/get attributes.

Although, we extensively and manually analyzed a large set of PRs for testability refactoring patterns, we neither expect our list to be complete nor final – and it should indeed be extended and refined further. However, what is particular in our contributed patterns is their empirical grounding based on a large dataset of GitHub PRs. The patterns with the description and examples can be used as a catalog for education and training or as "testability how-to" for contributor guidelines in community sites. The observed patterns are derived from multiple open-source projects, by multiple developers. They can be considered as common practices applied to production code in order to enhance unit testing. Typically internals of classes under test are exposed for control or assertion by unit tests or additional levels of indirection are introduced in order to substitute real components with subs.

The repeated **patterns occurrences** across various projects, developers, and PRs reveal a "common nature" of those patterns. However, the actual frequency of patterns may vary from project to project. It is important to note that the quantification of occurrences reported in this work cannot be generalized to the whole GitHub community since our samples were based on masks – except for the last 150 randomly selected Test-Pairs PRs in Table 1. Clearly, the specific studied samples M1-M9 represent a small fraction of all PRs (where we expected to find refactoring for testability). We might have missed other patterns and other occurrences in a full random sample.

Overall, we observed at least one of those patterns in about 25% of the analyzed Test-Pairs PRs. The analyzed PRs were sampled through the masks as we were looking to the patterns in the whole PR population. Therefore, the general frequency is likely to be smaller. The small analyzed random sample (mask Other) indicate that 13% of all test-pairs PRs might include one or more of the pattern. However, as PRs are less likely to focus on refactoring and testability as once, the share in other non-PRs commits might be higher. PRs *only including* unit tests changes and respective production code refactoring are especially interesting because they are isolated from new features or bugfixes in production code and can, e.g. be used to train repository mining tools. Generally, PRs with unit tests (and relevant testability refactoring to production code) only for the purpose of improving quality are less common, and might require a separate investigation. A representative quantification of the pattern frequencies can be achieved either with a large community effort or through automatically mining the patterns from source code changes.

In future work, projects with strict code coverage criteria may provide a good basis for additional analysis and mining of Test-Pairs PR, due to the expectation that production code is likely to be already tested prior to the creation of the PR or that testability refactoring is necessary for keeping the coverage level. Moreover, to assist **handling complex Test-Pairs PRs**, a repository mining tool can be implemented to automatically analyze PRs (and potentially commits), e.g. using java code parser such as INRIA Spoon to detect changes in production code, analyze usage of added methods/classes, and flag PRs with newly added methods used only in unit tests. One way of further research is to study how well our manually labeled dataset of PRs with the patterns can be used to train a classifier for deciding if a PR or another development situation is testabililty relevant or not.

The patterns "extract method for override" and "extract method for invocation" have been identified in 90 out of 230 testability refactorings investigated in this study and further analysis of metrics of the original methods can be useful to understand the optimal method length/complexity or other factors before it needs to be refactored. Clean code handbook [21] suggests smaller classes and shorter class methods, so it can be assumed that projects that follow Clean code require less refactoring for testability.

Finally, in our study we analyzed a number of **PRs with specific title masks** M4-M8, such as $singleton$ or $concurrency$. We explicitly considered those masks based on related work and our own experience, that, e.g., testing singletons or concurrency is rather difficult as well as evidence from the literature. However, we found that relatively few PRs contained similar problems. In one example pentaho-platform/1846 the "extract a method for override" pattern has been used to mock a singleton instance in SchedulerService::getSecurityHelper. More investigation into these specific design situations as singleton usage in open source projects would be useful in order to understand how such design situation actually affect testability. Instead of looking for PRs, e.g., with $singleton$ in the title, source code from various projects can be processed using a singleton-detector tool[8]. Similarly, the other masks can be studied separately using direct and likely more representative sampling from the source code.

In summary, our results can be used by practitioners, tool designers, and researchers. For practitioners we have created an online catalog of patterns[9], where they can find an appropriate testability refactoring pattern with a few examples, that can help to write unit-tests for a hard-to-test piece of code. Practitioners can contribute to the online catalog using standard github issues/PRs. The online catalog can also be used for education purposes, to teach refactoring techniques for legacy code, extending the work of Feather [18] with more examples from the open source domain.

Tool designers can benefit from the dataset[10] of manually reviewed testability refactoring PRs that can be used to recommend refactorings, prepare templates for refactoring patterns and tools to handle complex test-pairs PRs. Some testability refactoring patterns are easier to implement as a refactoring template: widen_access_for_invocation, override_system_time, extract_attribute_for_assertion require less knowledge about the context about the dependencies, while extract_class_for_invocation, extract_class_for_override, extract_method_for_override require more knowledge about side effects of the code and its dependencies.

Researchers can use the dataset for further research to better understand refactoring for testability, link code coverage to testability patterns, and specifically studying PRs with certain title/body masks.

---

[8] https://github.com/paul-hammant/java-singleton-detector
[9] https://github.com/icse2023preich/testability-refactoring-patterns/blob/main/catalog.md
[10] https://github.com/icse2023preich/testability-refactoring-patterns/blob/main/reviewed.csv

*B. Threats to Validity*

This study focused on popular open source projects hosted on GitHub and only projects in Java and JVM-based languages (occasionally Scala, Kotlin, Groovy) were selected. Projects developed in other programming languages, such as Python or Javascript, can be analyzed in a separate study. Due to the availability bias, it is easier to analyze open source projects. Future studies should thus target other types of projects and other ways of tracking changes.

All 724 PRs in this study were analyzed manually by a single person (the first author with extensive development experience). About a 60% of those PRs were additionally and independently analyzed by one of three other postgraduate students, each also with several years of development experience. We conducted the second additional independent labeling iteration as a quality assurance measure.

As every manual labeling work, our might still be error-prone. In order to mitigate this source of bias, we designed the manual analysis task so that every PR get reviewed at least twice: once to assess whether the change was related to testability and once more to identify the actual testability refactoring and categorize its corresponding pattern. To encourage replication, we share the dataset online. Since our patterns are clearly defined and the labeling of code changes provides less room for interpretation as, e.g., labeling natural language text, we think that our results have a solid reliability, even if the exact ratios of the occurrences might be slightly different. We think that the probability of missing testability patterns in PRs (i.e. false negative errors) is higher than falsely labeling a PR as relevant while it is not (false positive errors). Therefore, the actual prevalence of patterns can be higher.

We calculated the coder agreement rate (Cohen's [22] Kappa) to be ∼70% which is considered substantial. In the majority of PRs (258 out of 290 PRs), both reviewers found the same patterns (same matches). In 7 PRs at least one pattern was identified by both reviewers. In 56 PRs reviewers found different patterns or one of them considered the PR irrelevant for testability. When looking at and resolving the conflicting reviews, we found the reason for disagreement in almost all cases was because of the reviewer oversaw a pattern.

Finally, data collection, preparation and analysis is heavily based on analysis scripts which we developed for this study. The scripts might include unintentional defects that might bias some of the quantitative results. To reduce this risk, we reviewed the code and share it in our replication package. As the quantitative results were consistent and complemented the manual analysis, we think this threat is minimal.

## VII. RELATED WORK

We focus our related work discussion on testability, pull request mining, and co-evolution of production and test code.

*A. Testability*

There is a large body of knowledge on software testabililty. Garousi et al. [7] recently surveyed over 200 papers in the field. To the best of our knowledge, none of those mined open source repositories to systematically identify and quantify testability related refactoring. Garousi et al. found that only 12% of these papers presented empirical foundation studies to understand testability. Our study helps to fill this gap by contributing an empirically grounded list of ten higher-level testability refactoring patterns. The patterns help make members of classes visible and controllable by unit tests.

Sae et al. [23] investigated how developers prioritize code smells for refactoring. They observed that poor testability is one of the reasons to refactor production code, for instance, when a bugfix for a release requires a unit test and the underlying production code needs to be refactored first. This work is complementary to our since we focused on studying the actual changes and the refactoring rather than investigated the reasons for the refactoring. One follow up study could be to use our patterns and the quantitative and qualitative findings to, e.g., conduct an interview study to better understand developers behavior before and after conducting the refactoring for testability. Payne et al. [24] argued that object-oriented concepts such as information hiding and inheritance have a negative impact on easiness of testing. Some of the refactoring patterns identified in our paper such as widen_access_for_invocation essentially make class members visible for unit-tests.

In their empirical study, Bruntink and van Deursen [25] considered larger classes as harder to test. The authors observed that the Spearman's rank correlation between the number of lines of code in production (LOC) and test code (TLOC) is relatively high (0.5 or above). Badri et al. [26] studied LOC and Chidamber-Kemerer metrics of test-pairs [27]. Terragni et al. [28] also studied the relationship to code coverage. Moreover, Alshayeb et al. [29] conducted experiments with undergraduate and graduate students, who refactored the code of 3 different systems. Testability was measured using Chidamber-Kemerer metrics that Bruntink and van Deursen [25] consider to be good predictors of testability. All these studies motivated our work and complement it as we focus on identifying and characterizing a set of recurrent composite testability refactoring patterns through manual and automated analysis in a large pull request dataset. These studies with ours can help design tool support to assist refactoring for testability.

*B. Pull requests and GitHub mining*

Mining GitHub data in general and pull requests in particular have become a popular research topic in recent years. Gousios and Zaidman [30] shared a dataset for pull-based development research. In this study we have obtained PRs from the [12] dataset. Gousios et al. [31] explored pull-based software development in GitHub (fast turnaround, increased opportunities for community engagement, and decreased time to incorporate contributions) focusing on the decision to merge a pull requests and the time to process it. Eirini et al. [32] mined GitHub data, in particular pull requests and identified perils and promises. One of the promises is that interlinking of developers' data, pull requests, issues, and commits provides a comprehensive view of software development activities. This is one of the reason for us to first focus on pull request data in our study. Previous work on PR mining focuses on understanding the review process in open source context, particularly code merging activities and decision criteria to accept or reject pull requests.

Yu et al. [33] argued that "maintaining software quality seem to be one of top priorities of integrators" in pull based development. This also motivated our study and the data selection. Our research not only automatically analyzes a random sample of pull requests to study the prevalence of atomic refactorings. It also extensively reviewed several pull request samples to assess whether they are actually covering testability refactoring and what common patterns can be observed across the projects.

Considering the refactoring in pull requests, Coelho et al. [34] investigated PRs where refactorings have been suggested during a code review and implemented as separate commits. In our study, we focus on the test-pairs Pull Requests and on the testability relevant refactoring. We look at the entire change set of PRs without differentiating between initial and refactoring-inducing commits, which can be investigated in a separate study. Sousa et al. [35] introduce the idea of composite refactorings that consist of multiple atomic refactorings of the same or different types and that are considered as one change. Our testability refactoring patterns can be considered as composite refactorings. Through our automated mining and analysis of co-occurrences we pave the way towards automatically learning how the higher-level refactoring pattern are composed of atomic minable refactorings.

*C. Co-evolution of production and test code*

Zaidman et al. [36] conducted one of the first large studies on the co-evolution of test and production code. While this study certainly inspired our work, it focuses on a higher-level perspective than ours. Zaidman et al. aimed at identifying increased test-writing activity, detecting certain milestones as releases, and detecting testing strategies. Our analysis focuses on the production code and on a rather micro-perspective. We studied individual dependent changes to production and test code (see example in Figure 1) and identified statistically significant differences in the nature and frequency of production code changes when test code is changed in the same pull request.

Athanasiou et al. [37] investigated the relationship between the quality of test code such as density of assertions and completeness (code coverage of production code), effectiveness (ability of test code to detect bugs), and maintainability (ability of test code to be adjusted to changes in production code). Vonken et al. conducted [38] a controlled experiment with 42 participants to understand whether refactoring of production code is easier if supported by unit tests. They concluded that unit tests neither allow quicker refactoring nor produce higher code after refactoring. Van Deursen et al. [39] investigated refactoring of test code and identified bad smells and refactorings specific to test code. Our patterns affect the test code but primarily targets the production code. Kashiwa et al. [40] investigated how refactorings of production code break test code. Marsavina et al. [41] investigated the co-evolution of production and test code in 5 open source projects and identified 6 patterns of co-evolution, such as test code is created/updated together with changes in production code, which roughly correspond to the PR group incl_ref_for_test in our dataset.

## VIII. CONCLUSION

With the increased popularity and importance of automated quality assurance in software development, testability – and in particular unit-testability – is becoming a more and more important design goal than ever before. In this study we automatically and manually analyzed pull requests to explore and better understand how developers refactor production code to improve its testability. From our automated analysis, we learned that pull requests including test-pairs changes are generally more refactoring intensive than other pull request. Developers seem to frequently change method signatures (as name, parameter and return types) in such pull requests. From the manual analysis we identified in 184 out of 724 analyzed PRs ten well-formed and consistent testability refactoring. We also identified how typical refactoring operations on production code are logically combined into testability refactoring patterns. Unsurprisingly, methods and classes were frequently extracted to test a smaller piece of functionality or emulate behavior of dependencies. Future work should focus on automatically identifying such refactoring patterns and providing process and tool support for developers and tester to increase testability while keeping the coordination effort between testing and coding at a minimal required level.

## ACKNOWLEDGMENT

The authors thank the anonymous reviewers for their constructive comments and helpful remarks. We also thank Tim Pietz, Tim Puhlfürß and Vincent Raudszus for their support with the labeling our dataset.